\documentclass[aip, jcp, reprint, floatfix]{revtex4-1}

\usepackage{amsmath}
\usepackage{amsfonts}
\usepackage{dsfont}
\usepackage{xparse}
\usepackage{graphicx}
\usepackage{subfigure}
\usepackage{color}
\usepackage{hyperref}

\ExplSyntaxOn
\NewDocumentCommand{\mref}{m}{\quinn_mref:n {#1}}
\seq_new:N \l_quinn_mref_seq
\cs_new:Npn \quinn_mref:n #1
 {
  \seq_set_split:Nnn \l_quinn_mref_seq { , } { #1 }
  \seq_pop_right:NN \l_quinn_mref_seq \l_tmpa_tl
  ( 
  \seq_map_inline:Nn \l_quinn_mref_seq
    { \ref{##1},\nobreakspace } 
  \exp_args:NV \ref \l_tmpa_tl 
  ) 
 }
\ExplSyntaxOff

\newcommand{\NN}{\mathbb{N}}
\newcommand{\RR}{\mathbb{R}}

\newcommand{\calO}{\mathcal{O}}

\newcommand{\norm}[1]{\lvert {#1} \rvert}

\newcommand{\ind}[1]{{\mathds{1}_{{#1}}}}
\newcommand{\vphi}{\vartheta}

\draft 

\makeatletter
\def\@email#1#2{%
 \endgroup
 \patchcmd{\titleblock@produce}
  {\frontmatter@RRAPformat}
  {\frontmatter@RRAPformat{\produce@RRAP{*#1\href{mailto:#2}{#2}}}\frontmatter@RRAPformat}
  {}{}
}%
\makeatother

\begin{document}

\title{Partial versus total resetting for L\'evy flights in d dimensions: similarities and discrepancies}

\author{Costantino Di Bello}
\email{dibello@uni-potsdam.de}
\affiliation{University of Potsdam, Institute of Physics \& Astronomy, 14476 Potsdam, Germany}

\author{Aleksei Chechkin}
\affiliation{University of Potsdam, Institute of Physics \& Astronomy, 14476 Potsdam, Germany}
\affiliation{Faculty of Pure and Applied Mathematics, Wrocław University of Science and Technology, Wyb. Wyspiańskiego 27, 50-370 Wrocław, Poland}
\affiliation{German-Ukrainian Core of Excellence Max Planck Institute of Microstructure Physics, Weinberg 2, 06120 Halle (Saale), Germany}

\author{Tomasz Grzywny}
\affiliation{Faculty of Pure and Applied Mathematics, Wrocław University of Science and Technology, Wyb. Wyspiańskiego 27, 50-370 Wrocław, Poland}

\author{Zbigniew Palmowski}
\affiliation{Faculty of Pure and Applied Mathematics, Wrocław University of Science and Technology, Wyb. Wyspiańskiego 27, 50-370 Wrocław, Poland}

\author{Karol Szczypkowski}
\affiliation{Faculty of Pure and Applied Mathematics, Wrocław University of Science and Technology, Wyb. Wyspiańskiego 27, 50-370 Wrocław, Poland}
 
\author{Bartosz Trojan}
\affiliation{Faculty of Pure and Applied Mathematics, Wrocław University of Science and Technology, Wyb. Wyspiańskiego 27, 50-370 Wrocław, Poland}

\date{\today}

\begin{abstract}
    While stochastic resetting (or total resetting) is a less young and more established concept in stochastic processes, partial stochastic resetting (PSR) is a relatively new field. 
    PSR means that, at random moments in time, a stochastic process gets multiplied by a factor between $0$ and $1$, thus approaching but not reaching the resetting position.
    In this paper, we present new results on PSR highlighting the main similarities and discrepancies with total resetting. Specifically, we consider both symmetric $\alpha$-stable L\'evy processes (L\'evy flights) and Brownian motion with PSR in arbitrary $d$ dimensions. 
    We derive explicit expressions for the propagator and its stationary measure, and discuss in detail their asymptotic behavior.
    Interestingly, while approaching to stationarity, a dynamical phase transition occurs for the Brownian motion, but not for L\'evy flights. 
    We also analyze the behavior of the process around the resetting position and find significant differences between PSR and total resetting.
\end{abstract}


\maketitle

\textbf{Consider an insurance company. At a random moment, a customer incurs an accident and gets a refund from the insurance. From the company's point of view, refunding someone corresponds to an instantaneous sudden loss of money. If we consider that the company's total wealth is aleatory, we could mathematically model it as a stochastic process with random jumps. 
In physical literature, we refer to these kinds of models as {\em stochastic processes with resets}. 
There has been a great interest in resetting processes in the last fifteen years. In this paper, we will consider a peculiar model of stochastic resetting named {\em partial resetting}. 
With partial resetting, we mean that the instantaneous jump in the stochastic process is proportional to the current value of the process.
In the most studied total resetting mechanism, instead, the stochastic process gets instantaneously set to a specific value independently of any past value of the process.
In this paper, we want to study how this kind of mechanism affects the behavior of generic L\'evy flights in arbitrary $d$ dimensions. We obtained, both analytically and numerically, that total and partial resetting, although being equivalent under many aspects, show some completely different behaviors.}

\section{Introduction}
\label{sec:intro}
In the search for a hidden target, it is sometimes beneficial to restart from the beginning \cite{zecchina_prl}. This simple fact has inspired many researchers in the last decade. Although this idea can be found in some earlier works \cite{zecchina_prl, reset_1999}, we can with no doubt trace the birth of {\em stochastic resetting} with the seminal paper by Evans and Majumdar \cite{Evans}. This simple yet powerful idea of restarting has been the sparkle that started a long series of works, not only in the field of stochastic processes. Many applications were found in computer science \cite{restart_internet, rw_restart}, physics \cite{Sandev_2022}, finance \cite{jolakoski_chaos}, and biology \cite{edgar_pre}. 
It was found in \cite{Evans, Evans_maj_jpa, chechkin_sokolov_prl} that a random searcher moving with resetting takes a significantly lower time to reach its target. 
After less than 15 years it is already difficult to mention all the relevant papers in the field, and some topical reviews\cite{Evans_review, review_gupta, review_interact_res, Kundu_Reuveni_2024, Pal2024} are already available.
Recently, a new variant of the stochastic resetting model, sometimes called {\em partial stochastic resetting} \cite{marcus_psr, Pierce, shlomi_PSR, costantino, K_Olsen_PSR1, K_Olsen_PSR2, biroli_pre}, has been introduced in the physics literature. Considering a generic stochastic process $X_t$, partial stochastic resetting (PSR) consists of multiplying, at random times $T_n$, the current position of $X_{T_n}$ for a certain factor $c\in [0,1]$. 
This kind of model is not completely new: the {\em additive-increase multiplicative-decrease} (AIMD) model is the analogous one. AIMD was used to describe the transmission of data via internet using the transmission control protocol (TCP) \cite{dumas_AIMD_algo, guillemin_AIMD, Kemperman, vanderhofstad, lopker_j_app_prob, leeuwarden_TCP}. 
Nevertheless, the limit of AIMD is that it is a piecewise deterministic model, in the sense that the underlying process $X_t$, i.e. the process between consecutive partial resets, is deterministic.
New works on PSR \cite{costantino, Pierce, shlomi_PSR, K_Olsen_PSR1, K_Olsen_PSR2}, instead, take into account the stochasticity of $X_t$.
The peculiarity of PSR of having sudden instantaneous crashes makes it a suitable model for actuarial sciences\cite{ruin_probabilities}.\\
The scope of our paper is twofold. After a brief review of the known facts about PSR, we will first discuss new analytical results on multidimensional isotropic L\'evy flights with PSR. Second, we will present all the features of PSR making a comparison with total resetting. 
As analytical derivations are heavily technical and accessible only to an extremely specialized audience, we preferred reporting them in a companion paper\cite{my}. 
Here we focus on the results and not on the technicalities of the derivation.

\section{The model}

\begin{figure}[h]
    \centering
    \subfigure[]{\includegraphics[width=0.48\linewidth]{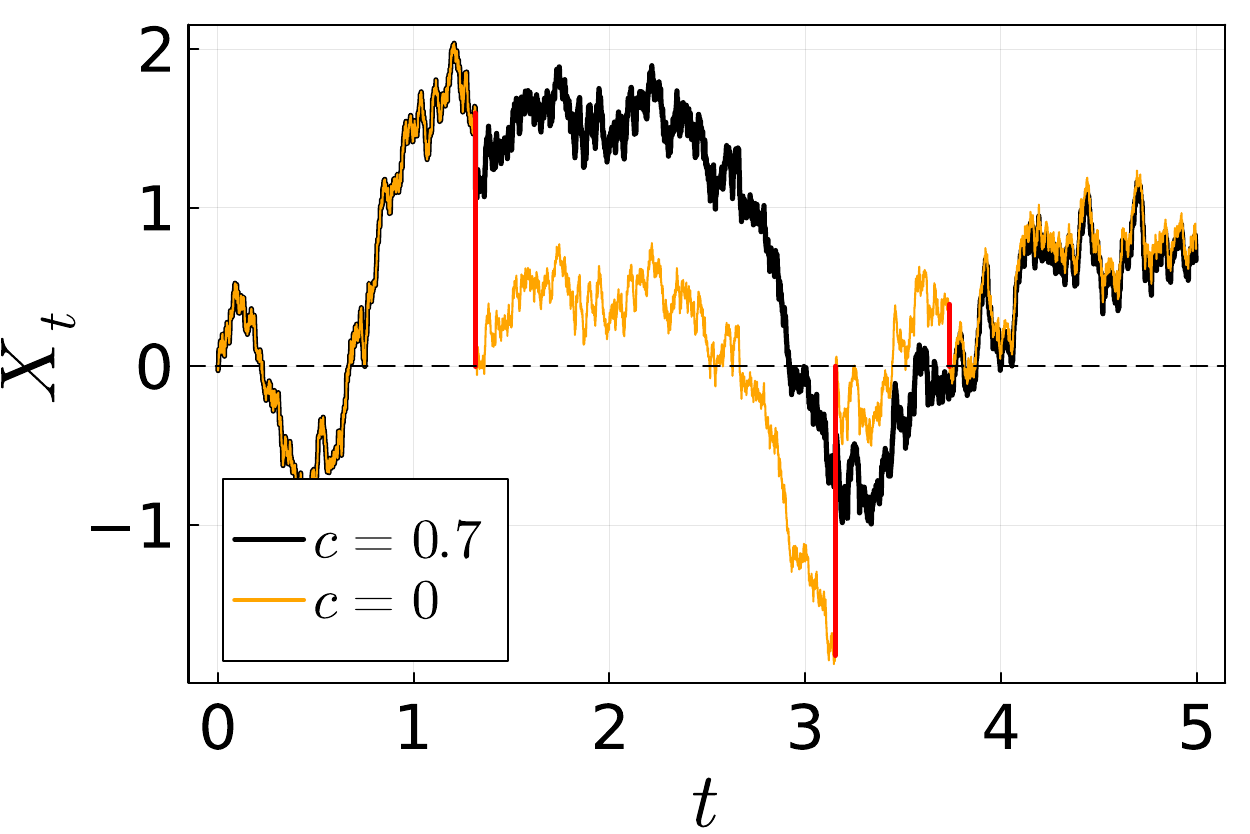}}
    \subfigure[]{\includegraphics[width=0.48\linewidth]{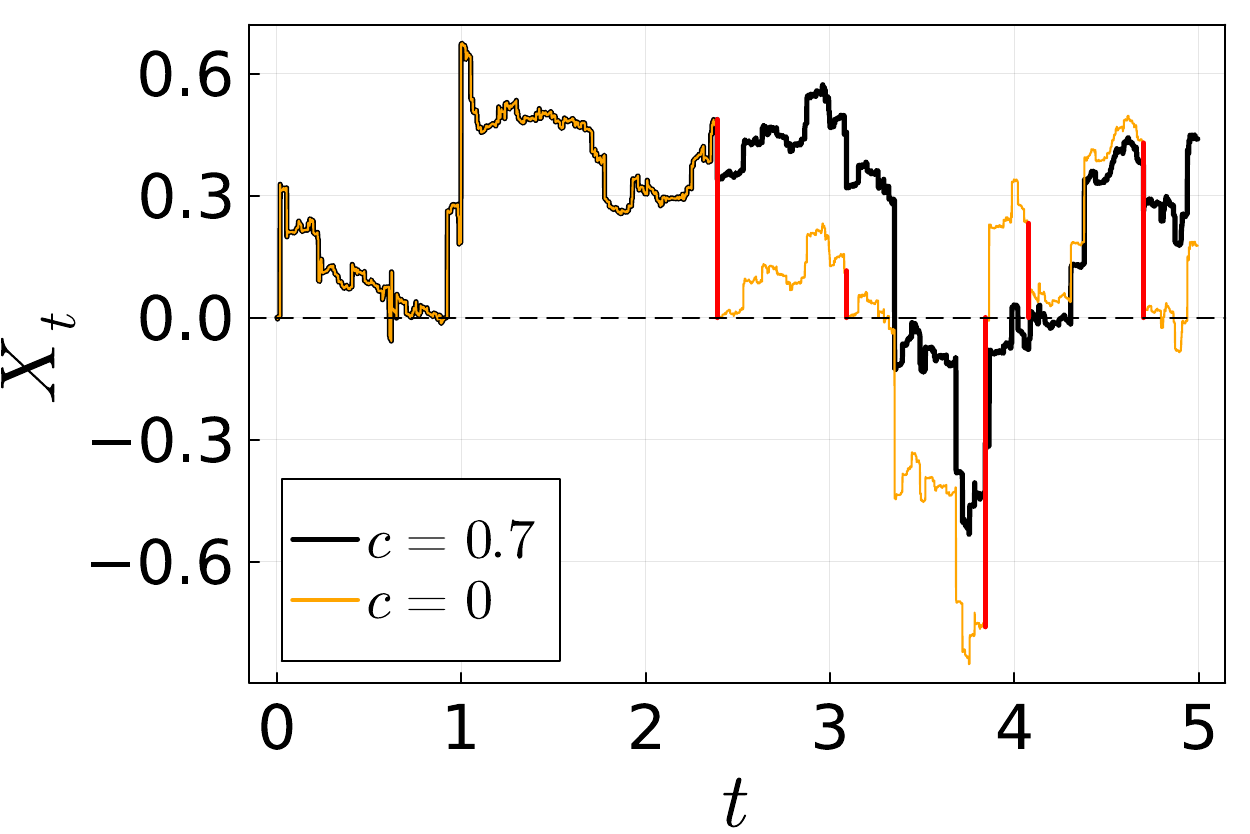}}
    \caption{Comparison between total resetting (yellow line) and partial resetting (black line). Red lines highlight resetting events. In panel (a) we set $\alpha=2$ (Brownian motion), while in panel (b) we set $\alpha =1$ (Cauchy).}
    \label{fig:trajectories}
\end{figure}

Following \cite{my}, we denote here with $Y_t$ d-dimensional isotropic L\'evy flights (LF) with stability index $\alpha \in (0,2]$, that is $\alpha$-stable L\'evy process \cite{Kyprianou_Pardo_2022}, while $X_t$ is the LF with PSR.
The case $\alpha=2$ refers to the Brownian motion. At random times $(T_i)_{i \in \mathbb{N}}$ the stochastic process $X_t$ gets multiplied by a factor $c\in [0,1]$, and we assume that inter-arrival times $T_i - T_{i-1}$ are independent and exponentially distributed with the same rate (inverse mean) $r$ $\forall i \in \mathbb{N}$. 
In other words, for each infinitesimal time interval $dt$ there is a probability $rdt$ that a partial resetting event occurs, moreover, between two inter-arrival times the increments of the two processes $X_t$ and $Y_t$ coincide.
Therefore, we can summarize the evolution of $X_t$ with the equation
\begin{equation}
    \label{eqn:evolution}
    X_t =
    \begin{cases}
        Y_t, & \text{if } t<T_1 , \\
        c X_{T_n^-} + Y_t - Y_{T_n}, & \text{for } t \in [T_n, T_{n+1}),\, n\in \mathbb{N}.
    \end{cases}
\end{equation}
A graphical representation of the evolution of the $X_t$ for total and partial resetting is available in figure \ref{fig:trajectories} for two different one-dimensional L\'evy flights.

The propagator of the process $Y_t$ with initial condition $Y_{t_0}=x_0$
is denoted by $p_0(x, t|x_0, t_0)$, where $x,x_0 \in \mathbb{R}^d$. Due to homogeneity in space and time of $Y_t$ we have
$p_0(x,t|x_0,t_0)=p_0(x-x_0,t-t_0|0,0) \equiv p_0(x-x_0,t-t_0)$. As it was shown in \cite{costantino}, $X_t$ is also homogeneous in time, therefore we can set $t_0=0$ everywhere without losing generality.
We recall that the propagator of a symmetric d-dimensional L\'evy flight, in terms of its Fourier transform, reads
\begin{equation}
    \hat{p}_0(k,t) \equiv \int_{\mathbb{R}^d} {\rm d}x e^{ik\cdot x} p_0(x,t) = \exp \left[i x_0 \kappa -D_\alpha |k|^\alpha t \right]
\end{equation}
where $\alpha\in ]0,2]$ is called stability index, while $D_\alpha$ is a scale parameter sometimes called {\em noise intensity}\cite{chechkin_review}, clearly $k\in \mathbb{R}^d$.
We are interested in the pdf $p_r(x,t|x_0)$ of the process $X_t$. An analytical, yet cumbersome, expression for $p_r$ in one dimension is already available in \cite{costantino}. Nevertheless, the authors of \cite{costantino}, apart from stating the analytical expression of $p_r$, did not discuss any feature of this propagator.
Here, we would like first to generalize the expression of $p_r$ to arbitrary d-dimensions, and second, to analyze some interesting features of this pdf like moments, tails, approach to stationarity, and non-equilibrium steady state. 
A complete proof of the results that we present here is available in the companion paper \cite{my}. We prefer to split the work into two papers for multiple reasons. First, the complete analytical proof of our results is extremely long and requires advanced mathematical techniques which might be of interest to an extremely specialized mathematical  audience, only. 
Second, we would like to present the main features of partial resetting, highlighting the similarities and discrepancies from total resetting. 
The paper is organized as follows. In section \ref{sec:prel} we recapitulate the mathematical description of the model and all the known results already available in the literature.
In section \ref{sec:splines} we present an alternative expression of the propagator obtained with the technique of splines. In section \ref{sec:stat} we discuss in full detail the stationary distribution, its moments and its scaling for large $|x|$ in full details. In section \ref{sec:NESS} we discuss the relaxation of the propagator to a stationary state and the emergence of a dynamical phase transition for Brownian motion. In section \ref{sec:boundedness} we analyze the behavior of the propagator around $|x|=0$.
In all the sections the features of partial resetting are compared to the ones of total resetting.
A complete summary of all the features discussed, compared to the ones of total resetting, is available in Table \ref{table}.

\section{Preliminaries}
\label{sec:prel}
\subsection{Useful notations}
Let us first give a few notations that we are going to use in the entire manuscript.
We denote with 
\begin{equation}
	\label{def:m}
	m \equiv c^\alpha.
\end{equation}
We will make extensive use of the $q$-Pochhammer symbols which are defined as
\begin{equation}
    \begin{aligned}
        (a; q)_0 \equiv  1,\qquad &
        (a; q)_n \equiv  \prod_{j = 0}^{n-1} (1-aq^j),\qquad \\
        (a; q)_\infty &\equiv  \prod_{j = 0}^\infty (1 - a q^j),
    \end{aligned}
\end{equation}
and $q$-Gamma function
\begin{equation}
    \Gamma_q(x) \equiv (1-q)^{1-x}\frac{(q;q)_{\infty}}{(q^x;q)_{\infty}}, \qquad x\notin -\mathbb{N}.
\end{equation}
\subsection{Known results}
Here, we recall some basics and known results on the PSR. Since in the whole manuscript we fix $x_0=0$, we will drop the dependence on $x_0$ in the propagator and we will write simply $p_r(x,t)$. The pantograph\cite{pantograph} Fokker-Planck equation (FPE) for this process in one dimension has been found in \cite{Pierce}. The generalization to many dimensions is straightforward and reads
\begin{equation}
    \label{eqn:FPE}
    \partial_t p_r(x,t) = D_\alpha \Delta^{\alpha/2} p_r(x,t) + \frac{r}{c} p_r\Big(\frac{x}{c}, t\Big) - rp_r(x,t).
\end{equation}
Here $\Delta^{\alpha/2}$ is the fractional Laplacian, defined in terms of its Fourier transform
\begin{equation}
    \mathcal{F} \left[ (-\Delta)^{\alpha/2} f(x) \right] (k) = |k|^\alpha \hat{f}(k),
\end{equation}
where $|k|$ is the euclidean norm of the vector $k \in \mathbb{R}^d$, while $\hat{f}(k)$ denotes the Fourier transform of $f$, defined as
\begin{equation}
    \hat{f}(k) = \int_{\mathbb{R}^d} {\rm d} x e^{ik\cdot x} f(x) .
\end{equation}
In the case of total resetting ($c \to 0^+$), the FPE reads (see \cite{res_d_dimensions})
\begin{equation}
    \partial_t p_r(x,t) = D_\alpha \Delta^{\alpha/2} p_r(x,t) + r \delta(x) - rp_r(x,t).
\end{equation}
It can be proved, using eq.~\eqref{eqn:FPE}, that for PSR the double Fourier-Laplace transform of the propagator, defined as 
\begin{equation}
    \hat{\tilde{p}}_r(k, s) \equiv \int_0^\infty e^{-st} \int_{\mathbb{R}^d} e^{ik\cdot x} p_r(x,t){\: \rm d} x {\: \rm d} t, \quad s>0, k \in \mathbb{R}^d,
\end{equation}
reads
\begin{equation}
    \label{Fpa}
    \hat{\tilde{p}}_r(k, s)=\sum_{n=0}^\infty r^n \prod_{j=0}^n \frac{1}{r+s+D_\alpha m^j |k|^\alpha}.
\end{equation}
Equation~\eqref{Fpa} is completely analogous to eq. (16) in \cite{costantino} for $d=1$. The only term depending on the dimension $d$ is the norm of the vector $|k|$. As expected the propagator is isotropic, as it only depends on $|k|$. \\
The stationary distribution $p^{(s)}(x)$ can be computed taking the limit $t\to \infty$ in eq.~\eqref{eqn:FPE}, thus it solves
\begin{equation}
    D_\alpha \Delta^{\alpha/2} p^{(s)}_r(x) + \frac{r}{c} p^{(s)}_r\Big(\frac{x}{c}, t\Big) - r p^{(s)}_r(x)=0,
\end{equation}
and its Fourier transform reads
\begin{equation}
    \label{Fpb}
    \hat{p}_r(k)=\prod_{j=0}^\infty \frac{r}{r+D_\alpha m^j |k|^\alpha}.
\end{equation}
Equation~\eqref{Fpb} is also a known result from \cite{costantino} (see eq. (27)) for the case $d=1$.\\
An analytical inversion of the double transform in eq.~\eqref{Fpa} has been perfomed in \cite{costantino} (see eq. (23)). The expression for $d$ dimensions reads
\begin{widetext}
    \begin{equation}
        p_r(x,t)= e^{-rt} \sum_{n=0}^{\infty} r^n\sum_{k=0}^n \dfrac{1}{(m^{-1}; m^{-1})_k(m;m)_{n-k}} \int_0^t\left((1-\delta_{n0})\dfrac{(t-t')^{n-1}}{(n-1)!} +\delta_{n0} \delta(t-t')\right)p_0(x,m^k t') {\: \rm d} t',
        \label{eqn:pr_Levy_flight}
    \end{equation}
\end{widetext}
where $\delta$ is the Dirac delta function (we used the same symbol for both discrete and continuous cases).
The propagator for total resetting has, instead, a much simpler expression \cite{res_d_dimensions}
\begin{equation}
    \label{eqn:renewal_total_res}
    p_r(x,t) = e^{-rt} p_0(x,t) + r\int_{0}^{t} e^{-rt'} p_0(x,t') {\rm d} t'.
\end{equation}
Formula \eqref{eqn:pr_Levy_flight} is rather cumbersome and difficult to handle. It is possible, in principle, to express the integral appearing in \eqref{eqn:pr_Levy_flight} in terms of hypergeometric functions (see eq. (36) in \cite{costantino}). Nevertheless, for an efficient numerical implementation, it is useful to use a different series representation, which we will discuss in the next section.
With such an efficient implementation, it will be possible to analyze the tails of the distribution with high accuracy and to verify the existence of a dynamical phase transition.\\
We conclude this section by mentioning the stationary distribution, also found in \cite{costantino} for one-dimensional processes (see eq. (32)).
It can be written in terms of Fox functions\cite{fox} and reads
\begin{multline}
\label{eqn:stat_Fox}
p^{(s)}_r(x)= \dfrac{1}{(c^\alpha;c^\alpha)_{\infty}} \sum_{n=0}^{\infty}\dfrac{1}{(c^{-\alpha};c^{-\alpha})_n}\frac{1}{\alpha|x|} \times \\
\times H^{2,1}_{2,3}\left[ \left(\frac{r}{D}\right)^{1/\alpha} \frac{|x|}{c^{n}}\left|\begin{array}{ll}(1,1/\alpha),(1,1/2)\\(1,1),(1,1/
\alpha),(1,1/2)\end{array}\right.\right].
\end{multline} 
In \cite{costantino} authors computed neither the moments nor the asymptotics. We will also expose our results for these two important features.\\
We do not report here the explicit formulae for total resetting, but they can be easily derived by taking the limit $c\to 0^+$ or $m \to 0^+$ (see references \cite{Evans, res_d_dimensions, kusmierz_pre}).

\section{Alternative representation of the propagator}
\label{sec:splines}
\begin{figure}[h]
    \centering
    \subfigure[]{\includegraphics[width=0.48\linewidth]{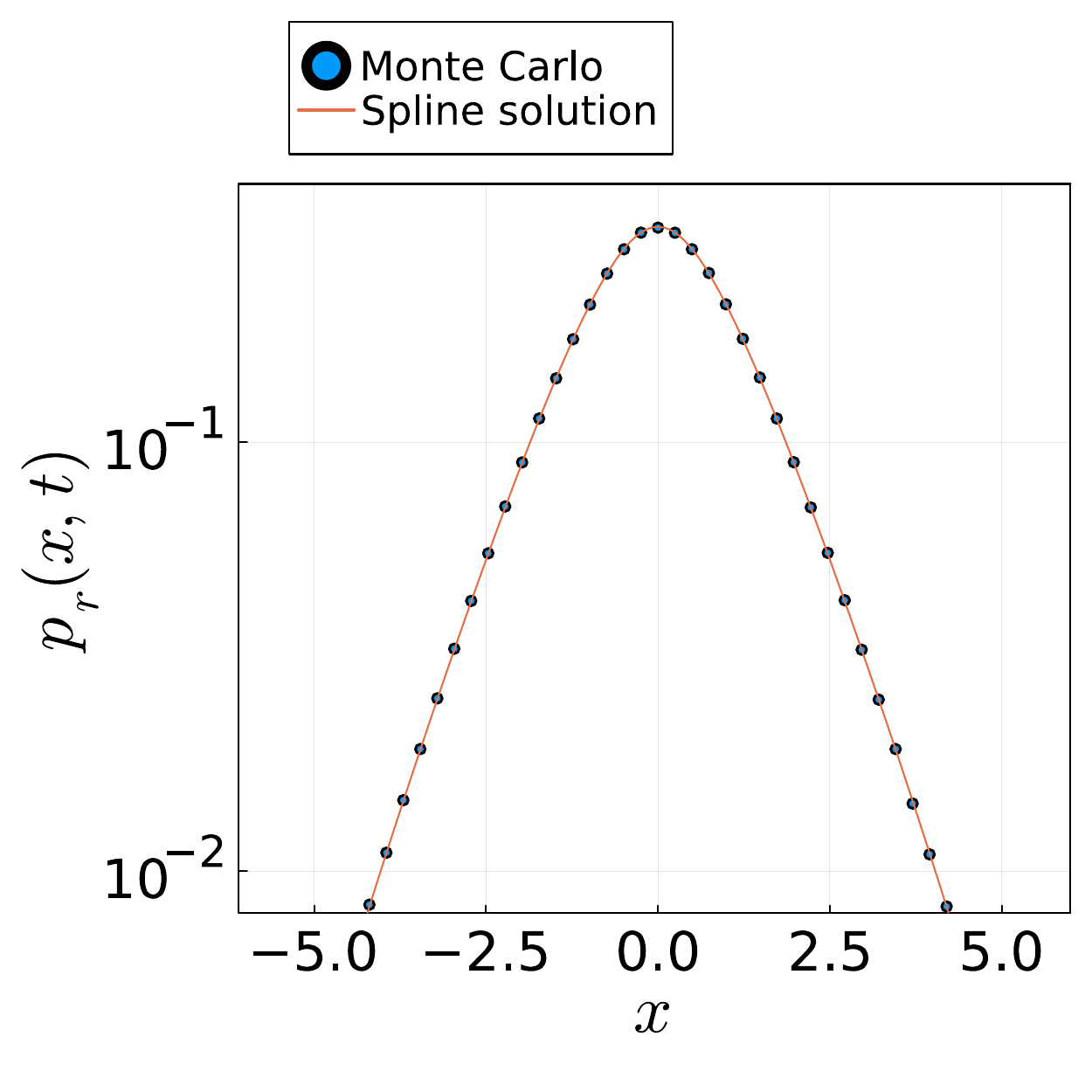}}
    \subfigure[]{\includegraphics[width=0.48\linewidth]{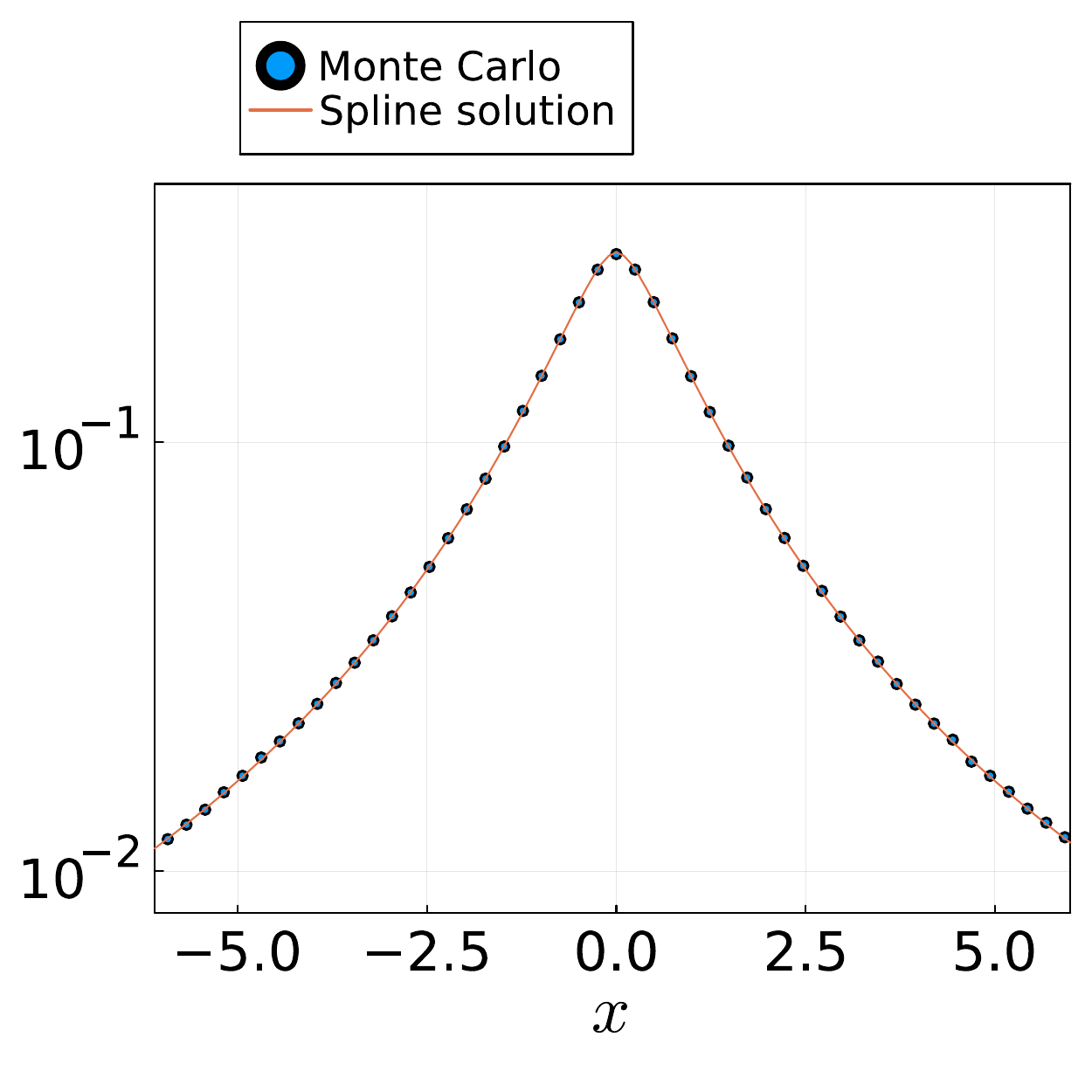}}
    \caption{Comparison in log scale between Monte Carlo sampling of positions (dots) with spline solution given in eqs.~\eqref{eqn:p_spline},\eqref{def:mu_t} for two different values of $\alpha$. In panel (a) we have $\alpha=2$, in (b) $\alpha=1$.  The Monte Carlo simulation samples $X_t$ using eq.\eqref{eqn:evolution}: we sampled resetting times $T_i$ and the increments between two resettings, thus the procedure does not suffer from the discretization of trajectories. Both histograms were made by generating $10^7$ values of $X_t$. Both simulations were made with parameters $d=1$, $D=1/2$, $t=5$, $r=1/2$, $c=0.5$.}
    \label{fig:simple_montecarlo}
\end{figure}

It is well known that the propagator $p_r(x,t)$ also solves the last renewal equation\cite{costantino}
\begin{multline}
    \label{keyidentity}
    p_r(y, t|x)
    =e^{-rt} p_0(y,t|x)
    + \\
    +r\int_0^t \int_{\RR^d}
    e^{-rs} p_0(z, s| x) p_r(y, t-s|cz) {\: \rm d} z {\: \rm d} s.
\end{multline}
The meaning of the rhs of~\eqref{keyidentity} is rather simple: either, with probability $e^{-rt}$, the random walk does not incur in any partial reset up to time $t$, and thus freely propagates from $x$ to $y$ with $p_0(y,t|x)$, or, with probability $re^{-rs}$, a partial resetting event occurs at some time $0<s<t$. 
This second possibility means that in the time interval $[0,s[$ the random walk propagates unperturbed from $x$ to some $z\in \mathbb{R}^d$ with propagator $p_0(z,t|x)$, then in the time interval $[s,t]$ the partial stochastic resetting process propagates from $cz$ to $y$ with $p_r(y, t-s|cz)$. 
Of course this term has to be integrated over all possible values of $s$ and $z$. 
The formula for total resetting~\eqref{eqn:renewal_total_res} can be retrieved by sending $c\to 0^+$ (see \cite{Evans}).\\
As shown in the companion paper\cite{my} (see section 3), a possible way to find the solution to the renewal equation is to define recursively the sequence $(P_n : n \in \NN)$ of splines via
\begin{align}
    P_1(u) &= \frac{1}{1-m} \ind{[m, 1]}(u), \label{eq:P1u}\\
    P_{n+1}(u) &= \max(u-m^{n+1}; 0)^n \int_u^1 \frac{P_n(v)}{(v-m^{n+1})^{n+1}} {\: \rm d}v, 
    \label{Pnu}
\end{align}
where $\ind{[m,1]}(u)$ is the indicator function of the interval $[m,1]$, i.e. it is $1$ if $u \in [m,1]$ and $0$ otherwise.
Note that each $P_n$ is supported on $[m^n, 1]$.
In \cite{my} it was proved that the solutions~\eqref{keyidentity} reads
\begin{equation}
    \label{eqn:p_spline}
    p_r(x,t)= \int_0^\infty p_0(x,u) \: \mu_t({\rm d} u),\quad x \in \RR^d,
\end{equation}
where $\mu_t$ is a measure defined as
\begin{equation}
    \label{def:mu_t}
    \mu_t({\rm d} u) = e^{-rt}\delta(t) {\rm d}u + e^{-rt} \sum_{j=1}^\infty r^jt^j P_j(u/t) \frac{{\rm d} u}{t}.
\end{equation}
We do not report here the proof, but just specify that it makes use of the renewal equation \eqref{keyidentity} and of the properties of $\alpha$-stable distributions (see \cite{sato} chapter 3). Notice that this solution is valid only for $c>0$. In the case $c=0$ (total resetting) the integral in eq.\eqref{eqn:p_spline} does not converge. As stated before, we do not report here the proof, but we mention that it makes use of the scaling properties of L\'evy flights\cite{sato}. This alternative representation of the propagator has several advantages. The first is that it is possible to analyze the tails of the distribution, and thus to identify, in the case of Brownian motion with PSR, a dynamical phase transition. 
Second, in this way it is possible to avoid using special functions that are generally not numerically stable. 
A comparison between a simple Monte Carlo simulation and the aforementioned spline solution is available in Fig.~\ref{fig:simple_montecarlo}.\\
In the next section we proceed with a discussion on the stationary distribution and its scaling for large $|x|$. 

\section{Stationary distribution}
\label{sec:stat}
As to the stationary distribution, one can prove, using the same spline method (see \cite{my} for details), that the stationary PDF $p^{(s)}_r$ for a d-dimensional process reads
\begin{multline}
\label{stationaryPDF}
    p^{(s)}_r(x)=
    \frac{1}{(m; m)_\infty}\sum_{k=0}^\infty (-1)^k \frac{m^{\frac{1}{2}k(k-1)}}{(m; m)_k} \times \\
    \times \int_0^\infty e^{-m^{-k} u} r^{d/\alpha}p_0(r^{1/\alpha}x,u) {\: \rm d}u,
\end{multline}
where $x\in \mathbb{R}^d$.
In the case of Brownian motion ($\alpha=2$) the integral can be computed explicitly,
\begin{multline}
    \label{stationaryPDFalpha2}
    p^{(s)}_r(x)= \frac{1}{(2\pi)^{d/2}}
    \frac{(r/D)^{d/2}}{(m; m)_\infty} \sum_{k=0}^\infty (-1)^k \frac{m^{\frac{1}{2}k(k-1)}}{(m; m)_k} \times \\
    \times \left(m^{\frac{k}{2}}\sqrt{\frac{r}{D}}|x|\right)^{1-d/2} K_{d/2-1}\left(\frac{\sqrt{r/D}|x|}{m^{k/2}}\right),
\end{multline}
where $K_{d/2-1}$ is the modified Bessel function of the second type, and we simply set $D_2=D$.
For $d=1$ in the case of Brownian motion we have
\begin{equation}
    \label{eqn:stat_BM_1d}
    p^{(s)}_r(x)= \frac12 \frac{\sqrt{r/D}}{(m; m)_\infty}\sum_{k=0}^\infty (-1)^k
    \frac{m^{\frac{k^2}{2}}}{(m; m)_k}
    e^{-m^{-\frac{k}2 }\sqrt{r/D}|x|}
\end{equation}
which is in agreement with equation (37) of \cite{costantino}, see also \cite{Pierce}.

\subsection{Asymptotic tail of the propagator and the stationary measure}
Recall that the L{\'e}vy measure of an isotropic $d$-dimensional L\'evy flight with index ${\alpha\in (0,2)}$ equals (see \cite{sato})
\begin{align}
    \label{eq:levymeasure}
    \nu(x)=\frac{2^{\alpha}\Gamma((d+\alpha)/2)}{\pi^{d/2}|\Gamma(-\alpha/2)|} |x|^{-d-\alpha}.
\end{align}
In \cite{my} it was proved that in the case of the L\'evy flight with index $0<\alpha<2$ we have
\begin{equation}\label{astalphaless2}
    \lim_{\substack{|x| \to +\infty \\ t \to+\infty} }
    \frac{p_r(x,t)}{\nu(x)} =\frac{1}{r(1-m)}
\end{equation}
and therefore
\begin{equation}
    \label{eqn:scaling_stat_LF}
    p^{(s)}_r(x)\sim \frac{1}{r(1-m)} \nu(x) \qquad \text{as $\norm{x}\to +\infty$.}
\end{equation}
where we write $f(x)\sim g(x)$ if $\lim_{|x|\to+\infty} f(x)/g(x)=1$.

The asymptotics \eqref{astalphaless2} is a consequence of one big jump principle: the large values of $|x|$ are possible only when we have one big jump with
regularly varying density \eqref{eq:levymeasure} that should happens till time $t$. \\
On the other side, if $Y_t$ is a Brownian motion in $\RR^d$, then from \eqref{stationaryPDFalpha2} it follows that
\begin{multline}
    \label{eqn:scaling_stat_BM}
    p^{(s)}_r(x)\sim \frac12 \frac{(r/D)^{(d+1)/4}}{(m; m)_\infty}  (2\pi)^{-\frac{d-1}{2}} \norm{x}^{-\frac{d-1}{2}}e^{-\sqrt{r/D}|x|} \\
    \text{as $\norm{x}\to +\infty$}.
\end{multline}
We can observe that from eqs.~\eqref{eqn:scaling_stat_LF} and \eqref{eqn:scaling_stat_BM} it is possible to obtain the scaling for total resetting by sending $m \to 0$. In both cases, partial and total resetting exhibit the same scaling for large $|x|$. 

\subsection{Moments of stationary distribution}
With a simple calculation, as usual available in \cite{my}, it can be proved that the fractional moments of the stationary distribution
for all $-1<\gamma<\alpha$ take the form
\begin{multline}
    \label{eq:stat:_moments_LF}
    \int_{\RR^d} |x|^{\gamma}p^{(s)}_r(x) {\: \rm d} x=\langle |X_\infty|^\gamma\rangle
    = \\
    = \frac{\Gamma(\gamma/\alpha+1)}{\Gamma_m(\gamma/\alpha+1)} (1-m)^{-\gamma/\alpha}\, r^{-\gamma/\alpha}\, \langle|Y_1|^\gamma \rangle.
\end{multline}
Equation~\eqref{eq:stat:_moments_LF} is one of the new results of this paper.   
Additionally, if $d = 1$ and $Y_t$ is a symmetric $\alpha$-stable process with $\alpha \in (0,2)$,
then for $-1<\gamma<\alpha$\cite{ralf_physreports2000},
\begin{equation}
    \langle |Y_1|^\gamma\rangle
    =\frac{2^\gamma \Gamma (\tfrac{1+\gamma}{2})\Gamma (1-\tfrac{\gamma}{\alpha})}{\sqrt{\pi}\,\Gamma(1-\tfrac{\gamma}{2})}.
\end{equation}
Similarly, if $d = 1$ and $Y_t$ is a Brownian motion $B_t$, then for $\gamma>-1$ one can see that
\begin{equation}
    \langle |B_1|^\gamma\rangle =\frac{2^{\gamma}\Gamma \big(\tfrac{1+\gamma}{2}\big)}{\sqrt{\pi}}.
\end{equation}
In particular, for $k \in \NN$ and one-dimensional Brownian motion,
\begin{equation}
    \langle X_\infty^{2k-1}\rangle=  \int_{\RR} x^{2k-1} p^{(s)}_r(x){\: \rm d} x=0, 
\end{equation}
and 
\begin{equation}
    \langle X_\infty^{2k}\rangle=\int_{\RR} x^{2k} p^{(s)}_r(x){\: \rm d}x= r^{-k}\frac{(2k)!}{(m;m)_k}.
\end{equation}
Hence in this case the variance of the stationary distribution $X_\infty$
equals
\begin{equation}\langle X_\infty^2\rangle=\frac{2}{r(1-m)}.\end{equation}
The same result for total resetting could be obtained by taking the limit $m\to 0^+$. The results are summarized in Table \ref{table}.

\section{Relaxation towards stationarity}
\label{sec:NESS}
\begin{widetext}
    
\begin{figure}[h]
    \centering
    \subfigure[]{\includegraphics[width=0.46\linewidth]{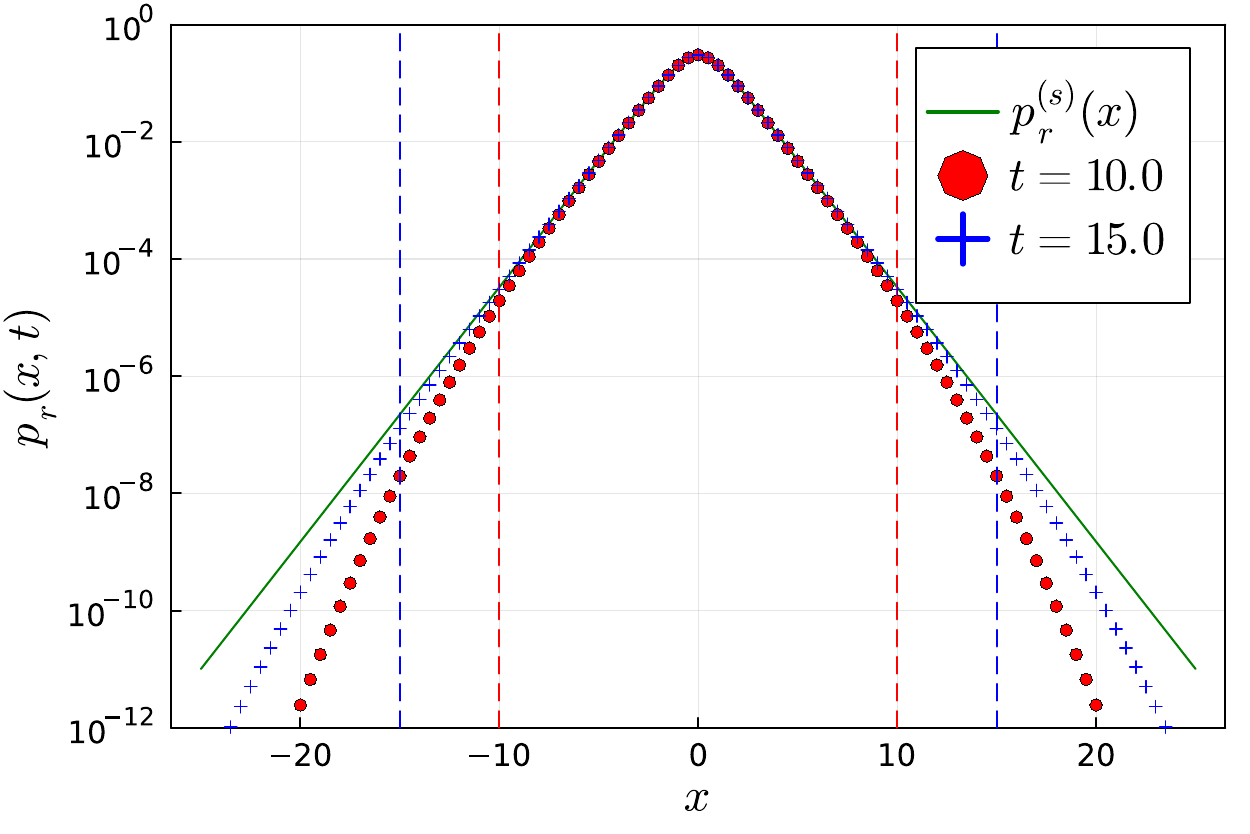}}
    \subfigure[]{\includegraphics[width=0.46\linewidth]{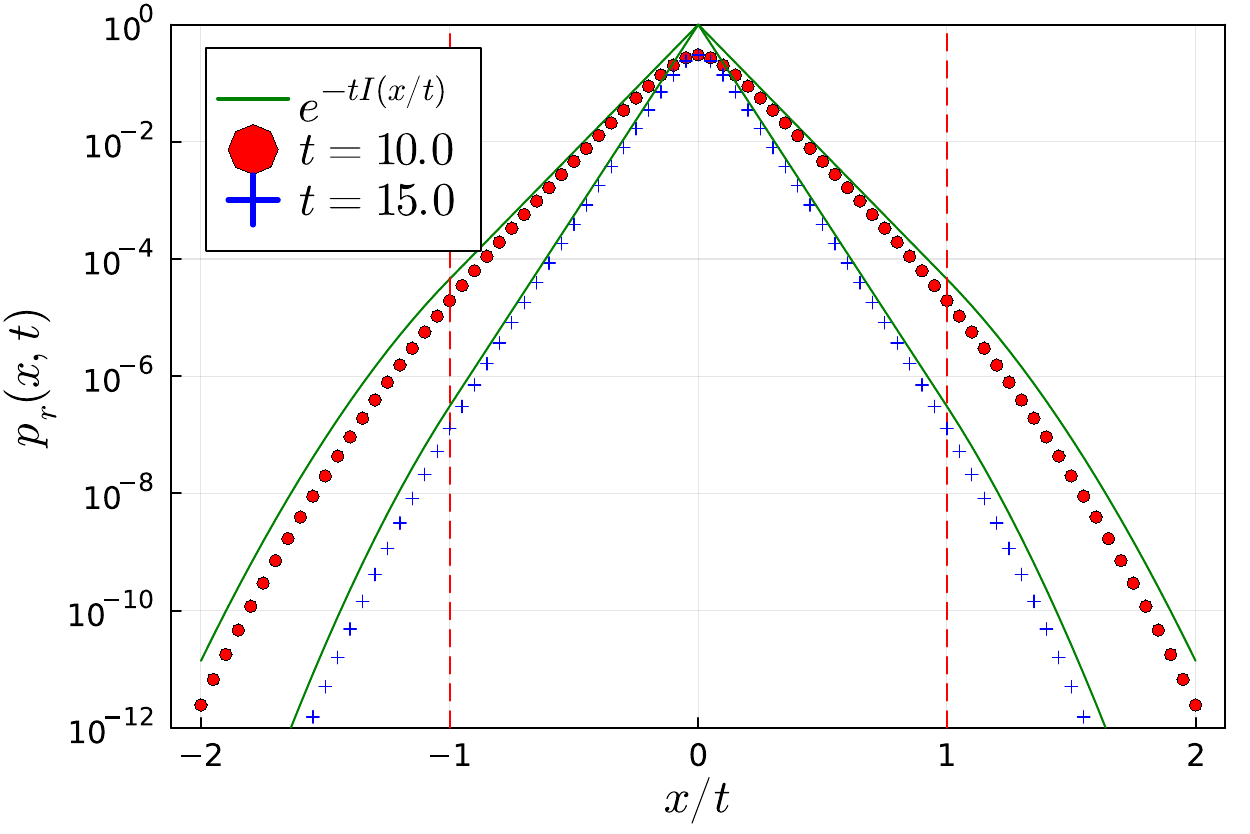}}
    \caption{Brownian motion with PSR in $d=1$, comparison between splines solution (dots) versus $p_r^{(s)}(x)$ (panel a) and $e^{-tI(x/t)}$ (panel b). In panel (a) the two vertical dashed lines separate the inner region, where stationarity has been achieved, from the transient region. In panel (b), close to $x=0$, the higher order exponentials in eq.~\eqref{eqn:stat_BM_1d} are non-negligible, and this explains the poor agreement for small $|x|$. Both figures were obtained with $D=0.5$, $c=0.5$, $r=0.5$.
    }
    \label{fig:brownian_prop}
\end{figure}

\end{widetext}

\begin{figure}[h]
    \centering
    \includegraphics[width=1\linewidth]{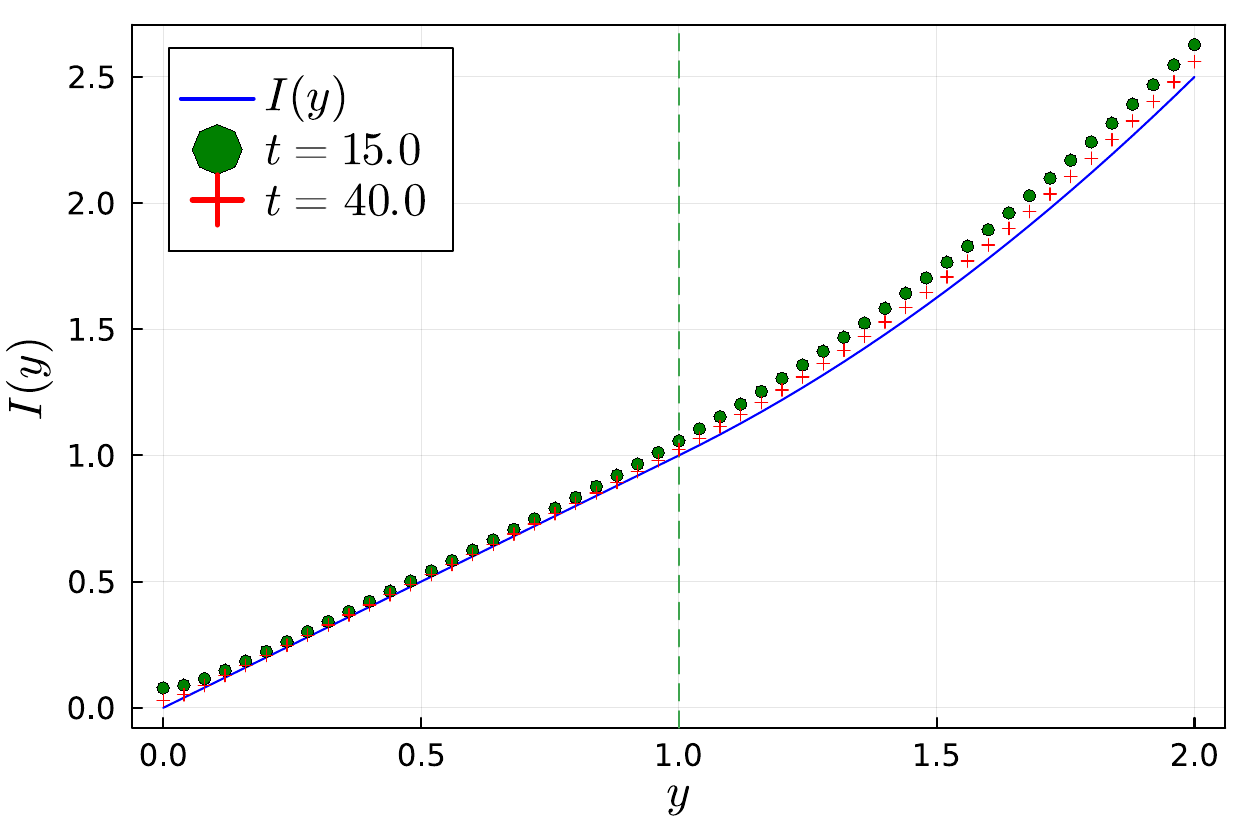}
    \caption{Large deviation function versus rescaled variable ${y=x/rt}$. All symbols represent the probability distribution for different values of time, the solid curve represents the large deviation function~\eqref{eqn:LDF}. The dashed line highlights the phase transition.}
    \label{fig:LDF}
\end{figure}

Partial resetting, similarly to total resetting, is characterized by the existence of a non-equilibrium steady state (NESS) and dynamical phase transition \cite{Touchette_dynphtrans}.
The concept is very well explained in \cite{maj_relaxation} for the case of Brownian motion in $d=1$ with total resetting. The idea is that when $t$ tends to $+\infty$ the propagator of the process can be written in the scaling form
\begin{equation}
    \label{eqn:scaling_general}
    p_r(x,t) \sim e^{-t I\left( x/t \right)}
\end{equation}
where the function $I$ is generally called {\em large deviation function} (LDF). If the LDF shows a discontinuity in one of its derivatives we are in the presence of a dynamical phase transition (we will come back later to the physical meaning of this kind of transition). In the literature on stochastic processes, the LDF is generally obtained either from the theory of large deviations or from saddle-point methods\cite{maj_relaxation}. 
In our companion paper \cite{my} we were able to prove that for $\alpha=2$ we have
\begin{widetext}
    \begin{equation}
        \label{eqn:prop_scale}
        p_r(x,t) =
        \begin{cases}
        p^{(s)}_r(x) \left(1 + \calO\left(t^{-1} \right)\right) & \text{ for } y<y^* \\
        e^{-rt}
        (4\pi Dt)^{-\frac{d}{2}} e^{-\frac{|x|^2}{4Dt}} \left[1 + \frac{4Dt^2}{\norm{x}^2}
        \vphi\left(\frac{4Dt^2}{\norm{x}^2}\right)+
        \calO\left(\frac{4Dt^2}{\norm{x}^2}\right)
        \right]  & \text{ for } y>y^*,
        \end{cases}
    \end{equation}
\end{widetext}
where $y=\frac{|x|}{t}$ is constant, $y^*=2\sqrt{Dr}$, $x\in \mathbb{R}^d$ and the function $\vphi(z)$ is defined as
\begin{equation}
    \vphi(z) = \sum_{j=0}^{\infty} \dfrac{1}{(m;m)_{j+1}} z^j, \quad |z|<1.
\end{equation}
This suggests that the LDF in eq.~\eqref{eqn:scaling_general} can be written as
\begin{equation}
    \label{eqn:LDF}
    I(y) = 
    \begin{cases}
        \sqrt{\dfrac{r}{D}} |y| & \text{for } |y|<y^*\\
        r+ \dfrac{y^2}{4D} & \text{for } |y|>y^*,
    \end{cases}
\end{equation}
which is precisely the same LDF obtained in \cite{maj_relaxation} for the total resetting. 
Interestingly, the dynamical phase transition is not affected either by the dimension or by the value $c \in [0,1[$. From this point of view, partial and total resettings are identical. A graphical representation of this result is presented in Figs.~\ref{fig:brownian_prop} and \ref{fig:LDF}. 
The result is proved in~\cite{my} by making use of \eqref{eqn:p_spline} and of the Stein method \cite{stein}. 
The physical meaning of this transition is the following: 
first of all, the propagator $p_r(x,t)$ can be interpreted as $\langle \delta \left( X_t-x \right) \rangle$, i.e. we could think of starting $N$ independent random walks and computing the histogram of positions at time $t$.
Second, since in the outer region $|x|>\sqrt{4Dr} t$ the far asymptotics of the propagator is Gaussian, the trajectories that never experienced resetting will be the ones contributing to the density $p_r(x,t)$ of this region. 
Therefore, following \cite{maj_relaxation}, we can argue that in the thermodynamical limit $N\to \infty$ the phase transition corresponds to a separation between the trajectories that reset many times with respect to the ones that never reset.

\begin{figure}[h]
    \centering
    \subfigure[]{\includegraphics[width=1\linewidth]{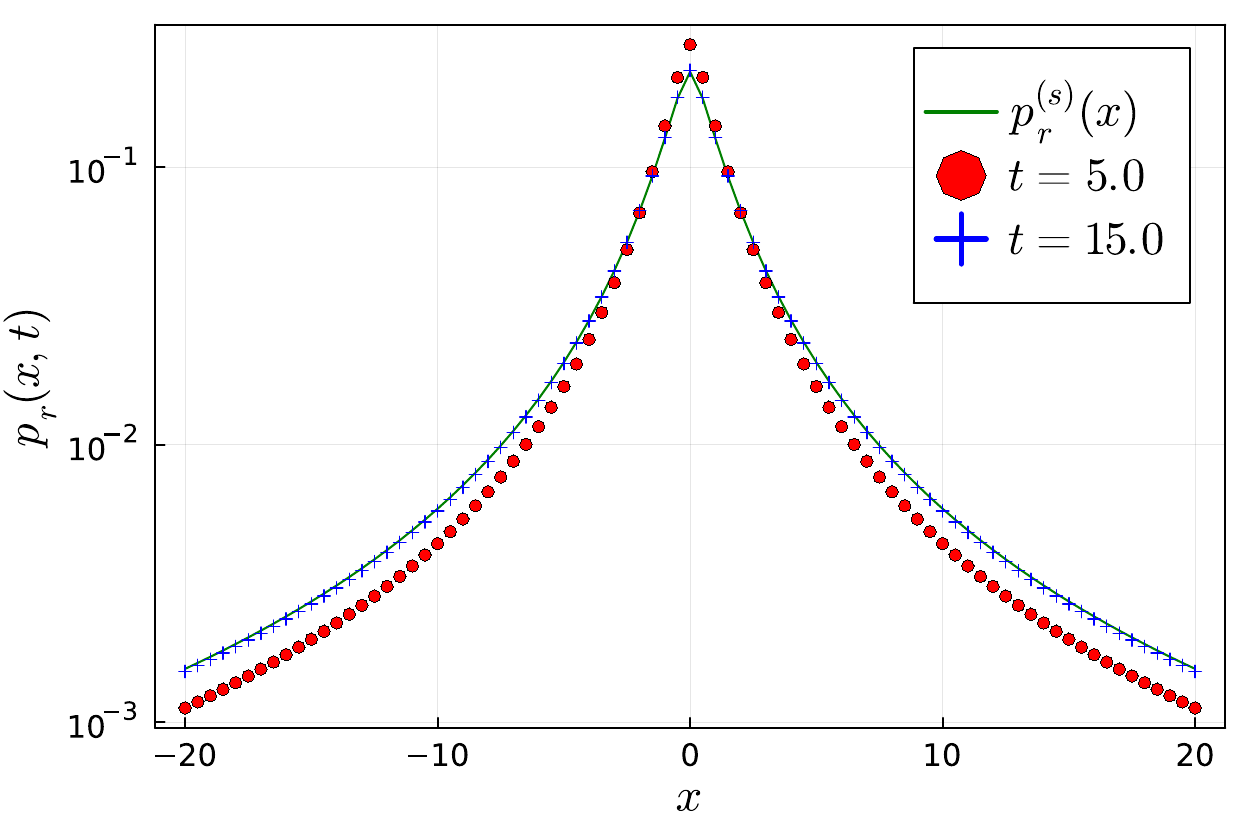}}
    \subfigure[]{\includegraphics[width=1\linewidth]{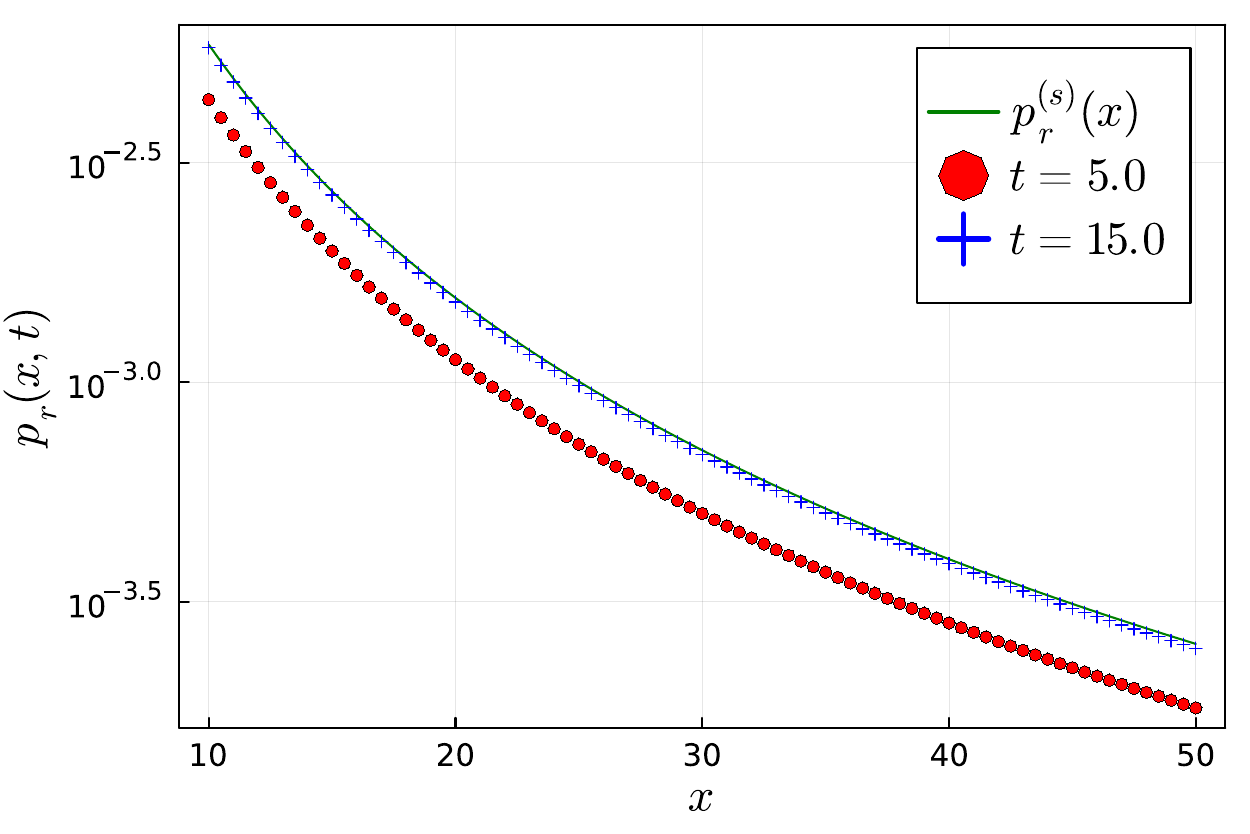}}
    \caption{Propagator of L\'evy flight with PSR in the case $\alpha=1$ (Cauchy). The symbols represent different values of time, while the solid line represents the stationary distribution given by eq.~\eqref{stationaryPDF}. Panel (b) shows a zoom of the far asymptotics of the distribution of panel (a).} The parameters of the simulations are $D=0.5$, $r=0.5$, $c=0.5$.
    \label{fig:stationary_LF}
\end{figure}

The dynamical phase transition is, instead, absent in the case of $\alpha\neq 2$. This is evident in Figure \ref{fig:stationary_LF}. An intuitive explanation for this behavior is the following. 
In Ref.\cite{Blumenthal1960} it was shown that the propagator of a LF for any $t>0$ and large $x$ scales as
\begin{equation}
    p_0(x,t)\sim t |x|^{-d-\alpha}
\end{equation}
Therefore, substituting the previous equation into \eqref{eqn:p_spline} we have that
\begin{equation}
    p_r(x,t)\sim A(t) |x|^{-d-\alpha}
\end{equation}
where $A(t)$ is some function of time such that $\lim_{t \to \infty} A(t) = {\rm const}$.
Moreover, the stationary distribution exhibits the same decay (compare eq.~\eqref{eqn:scaling_stat_LF}). 
Therefore, in the process of relaxation, the asymptotics of the propagator does not change, and a dynamical phase transition is absent for any $d$. 
The same argument applies in the case of LF with total resetting. Previous works, indeed, did not observe such a dynamical phase transition in $d=1$ \cite{kusmierz_prl, kusmierz_pre}.

\section{Boundedness of the propagator}
\label{sec:boundedness}
\begin{widetext}

\begin{figure}[h]
    \centering
    \subfigure[]{\includegraphics[width=0.46\linewidth]{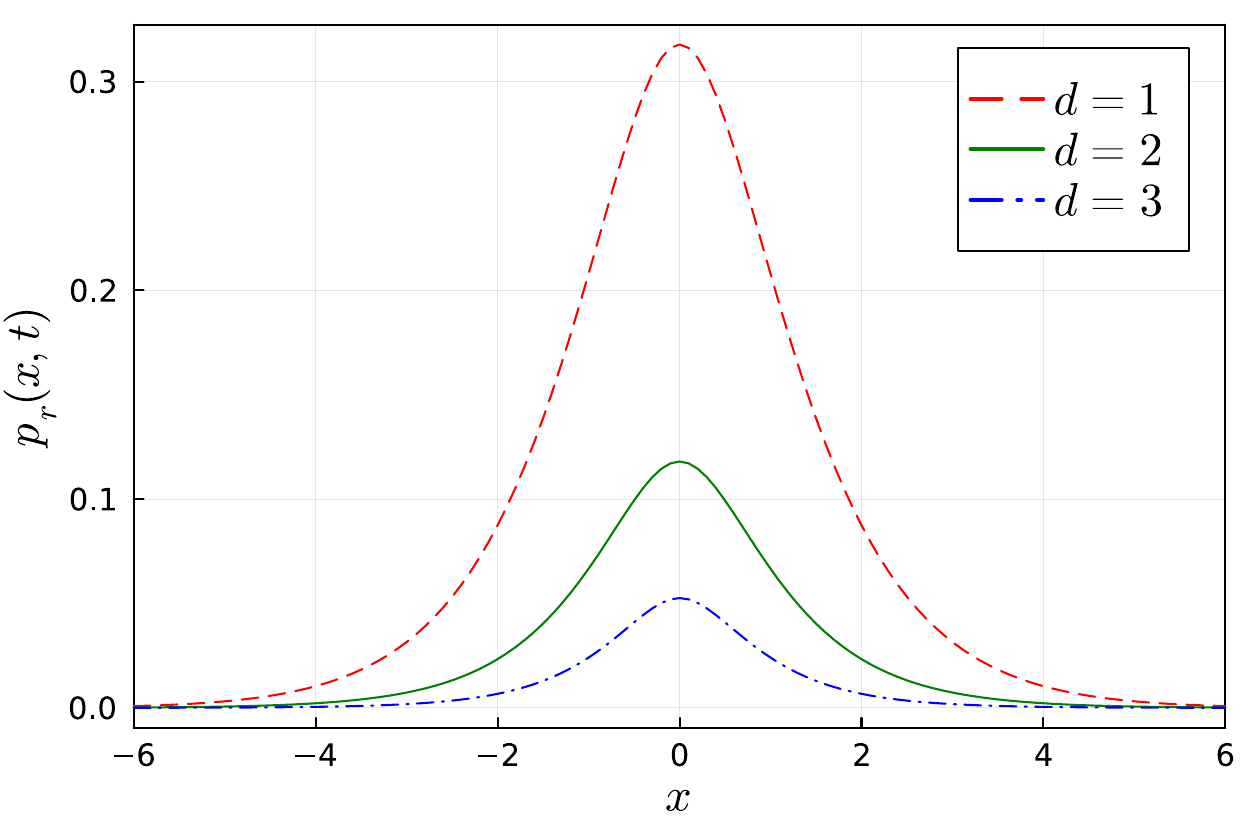}}
    \subfigure[]{\includegraphics[width=0.46\linewidth]{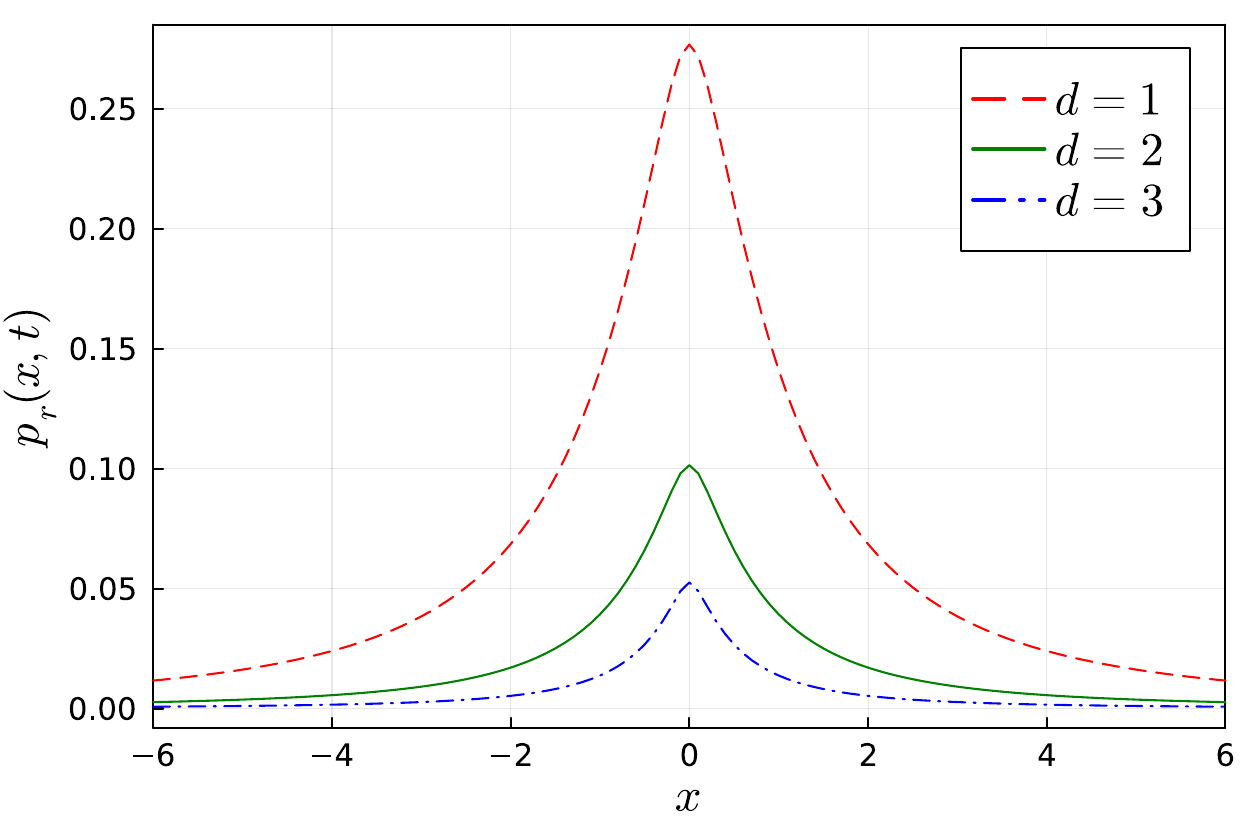}}
    \caption{Figure showing boundedness of the propagator. The parameters of the simulations are $D=0.5$, $r=0.5$, $c=0.5$, $t=5$, $\alpha=2$ (panel (a)), $\alpha=1$ (panel (b)). The plot represents just a ``slice" of the propagator, e.g. in the three-dimensional case we fixed $y=z=0$ and plotted with respect to $x$.}
    \label{fig:boundedness}
\end{figure}

\end{widetext}

As we have seen so far, there are no major differences between partial and total resetting. Scalings, tails, dynamical phase transitions are precisely the same for the two processes. Yet there is a feature, that we discuss in this section, that distinguishes between the two stochastic processes.
It has been observed in \cite{res_d_dimensions} that the propagator of a Brownian motion with total resetting in $d\geq 2$ dimensions is unbounded around $|x|=0$, i.e. ${\lim_{|x|\to 0} p_r(x,t) = +\infty}$. The reason for this fact is rather easy: the propagator of a LF has the scaling property
\begin{equation}
    \label{eqn:scaling_LF}
    p_0(x,t) = t^{-d/\alpha} p_0(x/t^{-1/\alpha}, 1)
\end{equation}
which, in the case of total resetting, after substituting it into eq.~\eqref{eqn:renewal_total_res} and taking the limit $|x|\to 0$ gives
\begin{multline}
    \lim_{|x|\to 0} p_r(x,t) = \lim_{|x|\to 0} \left[ e^{-rt} p_0(x,t) + \right. \\
    + \left. r\int_{0}^{t} e^{-ru} u^{-d/\alpha} p_0(x/u^{-1/\alpha}, 1 ) {\rm d} u\right],
\end{multline}
where the integral at the rhs diverges for $|x|\to 0$ whenever $d\geq \alpha$. On the other hand, for PSR after substituting the same scaling~\eqref{eqn:scaling_LF} into eq.~\eqref{eqn:p_spline}
we get
\begin{multline}
    \lim_{|x| \to 0} p_r(x,t) = \lim_{|x| \to 0} \left[ \int_{0}^{\infty} u^{-d/\alpha} p_0(x/u^{-1/\alpha}, 1 ) \mu_t(u) {\rm d}u \right] = \\
    = p_0(0,1) \int_{0}^{\infty} u^{-d/\alpha} \mu_t(u) {\rm d} u
\end{multline}
where the integral in the last step does not diverge since all splines are identically equal to $0$ close to $u=0$ (see eqs.~\eqref{Pnu}). In Fig.~\ref{fig:boundedness} we plotted the propagator for several values of $d$ and $\alpha$.

\section{Conclusions}
We have studied the behavior of partial stochastic resetting for symmetric L\'evy flights in arbitrary space dimension and compare it with total resetting. Interestingly, the two processes both show the existence of a stationary distribution with the same asymptotics, the same approach to stationarity and the same phase transition in the case of Brownian motion. The only feature distinguishing the two processes is the behavior around $|x|=0$. We have summarized all these features in Table \ref{table}.\\
We hope our paper will open many perspectives around
PSR. Future directions of research could be the study of
the first-passage time distribution for L\'evy flights\cite{koren_leapover_LF} with
PSR, or the extension of our theory to asymmetric L\'evy
flights\cite{padash_search_LF} and to non-stable L\'evy processes\cite{sato}, or even the case of PSR with cost\cite{cost_resetting} associated with each resetting.
Another 
interesting direction of the research could be studying the universal singularities of the models with resetting \cite{stella_pre,stella_prl}. We believe
that our work can serve as a model for practical applications, 
e.g. in finance or in actuarial sciences.

\begin{widetext}

    \begin{table}[h]
    \caption{\label{table} Comparison of $\alpha$-stable processes, $c\in [0,1[$}
    \resizebox{\textwidth}{!}{
    \begin{tabular}{ |p{3cm}||p{5.6cm}|p{5.6cm}|  }
     \hline
     Process $\alpha\in ]0,2]$ & $\alpha$-stable with PSR ($c \neq 0$) &$\alpha$-stable with total resetting ($c=0$)\\
     \hline
     FPE:  $\partial_t p_r(x,t) = $ & $D_{\alpha} \Delta^{\alpha/2}p_r(x,t) + \frac{r}{c} p_r\left( \frac{r}{c},t\right) - r p_r(x,t)$    & $D_{\alpha} \Delta^{\alpha/2}p_r(x,t) + r \delta(x) - r p_r(x,t)$ \\
     $\hat{\tilde{p}}_r(k,s)=$ &   $\sum_{n=0}^\infty r^n \prod_{j=0}^n \frac{1}{r+s+|k|^\alpha c^{\alpha j}}$  & $\frac{r+s}{s} \frac{1}{r+s+|k|^{\alpha}}$   \\
     $p_r(x,t)=$ & See eq.~\eqref{eqn:pr_Levy_flight} & $\lim_{c\to 0}$ in eq.~\eqref{eqn:pr_Levy_flight}\\
     Spline $p_r(x,t)$    & See eqs.~\mref{eq:P1u, Pnu, eqn:p_spline, def:mu_t} & Not available\\
     $\alpha \neq 2$: $p_r (x, t)\sim $&  $A(t) |x|^{-d-\alpha}$  & $A(t) |x|^{-d-\alpha}$\\
     & where $A(t) \to {\rm const}$ when $t\to \infty$ & \\
     $\alpha=2$: $p_r(x,t) \sim $ & $e^{-tI(x/t)}$ & $e^{-tI(x/t)}$\\
          &  with $I(y) = 
          \begin{cases}
              \sqrt{\frac{r}{D}} |y| \quad \text{if } |y|<y^*\\
              r+\frac{y^2}{4D} \quad \text{if } |y|>y^*
          \end{cases}$     &        \\
    $\langle |x|^\gamma p_r^{(s)}(x) \rangle = $  &  $\frac{\Gamma(\gamma/\alpha +1)}{\Gamma_m(\gamma/\alpha +1)} (1-m)^{\gamma/\alpha} r^{-\gamma/\alpha} \langle Y_1^\gamma \rangle $  & $\Gamma(\gamma/\alpha +1) r^{-\gamma/\alpha} \langle Y_1^\gamma \rangle $
    \\
       & with  $\langle Y_1^\gamma \rangle = \frac{2^\gamma \Gamma \left(\tfrac{1+\gamma}{2}\right)\Gamma \left(1-\tfrac{\gamma}{\alpha}\right)}{\sqrt{\pi}\,\Gamma\left(1-\tfrac{\gamma}{2}\right)}$ &   \\
     Boundedness & $p_r(x,t)$ bounded around $x=0$ & $p_r(x,t)$ {\bf not} bounded around $x=0$\\
     \hline
    \end{tabular}
    }
    \end{table}
    
\end{widetext}

\clearpage
\begin{acknowledgments}
The authors thank Dr Amin Padash for useful discussions.
This research is partially supported by the Polish National Science Centre under the grant No. 2023/49/B/ST1/00678. Aleksei Chechkin acknowledges support from the BMBF Project 01DK24006 PLASMA-SPIN-ENERGY.
\end{acknowledgments}

\bibliography{0mybib.bib}

\end{document}